\documentclass[extra,mreferee]{gji}

\usepackage[pdftex]{graphicx,color}
\usepackage{subfigure}
\usepackage{textpos}
\usepackage{gensymb}
\usepackage{hyperref}
\usepackage{chngcntr}
\usepackage{gji_extra}

\usepackage[latin1]{inputenc}
\usepackage{amsmath, amstext, amsfonts}
\usepackage{array, colortbl}

\makeatletter
\let\@tabclassz =\@classz
\makeatother 

\usepackage{todonotes}
\usepackage{url}

\newcommand{\aspect}{\textsc{Aspect}}

\definecolor{Gray}{gray}{0.9}
\newcommand{\highlight}[1]{%
  \fcolorbox{black}{gray!15}{$\displaystyle#1$}}

\usepackage{timet}

\title[A new formulation for coupled magma/mantle dynamics]
      {A new formulation for coupled magma/mantle dynamics}
\author[J. Dannberg, R. Gassm\"{o}ller, R. Grove, T. Heister]
{
Juliane Dannberg$^1$, 
Rene Gassm\"{o}ller$^1$,
Ryan Grove$^2$,
Timo Heister$^3$ \\
\small $^{1}$ \it University of California, Davis;
{\small \tt judannberg@gmail.com} and  {\small \tt rene.gassmoeller@mailbox.org}\\
\small $^{2}$ \it Mathematical Sciences, Clemson University; {\small \tt rrgrove6@gmail.com} \\
\small $^{3}$ \it Scientific Computing and Imaging Institute, University of Utah;
{\small \tt heister@sci.utah.edu}
}

\begin{document}

\label{firstpage}

\maketitle

\begin{summary}
Many open problems in the Earth sciences can only be understood by modelling the porous flow of melt through a viscously deforming solid rock matrix. 
However, the system of equations describing this process becomes mathematically degenerate in the limit of vanishing melt fraction. 
Numerical methods that do not consider this degeneracy or avoid it solely by regularising specific material properties 
generally become computationally expensive as soon as the melt fraction approaches zero in some part of the domain.

Here, we present a new formulation of the equations for coupled magma/mantle dynamics that addresses this problem, 
and allows it to accurately compute large-scale 3-D magma/mantle dynamics simulations with extensive regions of zero melt fraction. 
We achieve this by rescaling one of the solution variables, the compaction pressure, 
which ensures that for vanishing melt fraction, the equation causing the degeneracy becomes an identity and the other two equations revert to the Stokes system. 
This allows us to split the domain into two parts: 
In mesh cells where melt is present, we solve the coupled system of magma/mantle dynamics. 
In cells without melt, we solve the Stokes system as it is done for mantle convection without melt transport and constrain the remaining degrees of freedom. 

We have implemented this formulation in the open source geodynamic modelling code \aspect{} and illustrate the improved performance compared to the previous three-field formulation, showing numerically that the new formulation is optimal in terms of problem size and only minimally sensitive to model parameters. Beyond that, we demonstrate the applicability to realistic problems by showing large-scale 2-D and 3-D models of mid-ocean ridges with complex rheology.
Hence, we believe that our new formulation and its implementation in \aspect{} will prove a valuable tool for studying the interaction of melt segregating through and interacting with a solid host rock in the Earth and other planetary bodies using high-resolution, three-dimensional simulations.

\end{summary}


\maketitle

\section{Introduction}

Many Earth system processes are controlled by the porous flow of melt through a viscously deforming solid rock matrix. 
The equations that describe this process have been derived a long time ago \citep[e.g.][]{McK1984}: 
The motion of the solid is governed by Stokes flow, and the melt is transported according to Darcy's law. 
A large number of numerical models have been formulated that use these equations for different application cases
\citep[e.g.][]{Kat06, Katz2008, Weatherley2012, KMK2013, turner2017magmatic, butler2017shear, keller2017volatiles, katz2017subduction}. 
However, many formulations of coupled Stokes/Darcy flow break down in the limit of vanishing melt fraction (or porosity) 
because for this case the system is mathematically degenerate \citep{arbogast2017mixed}.

A common solution for this problem is introducing a cutoff or regularization for certain material properties or solution variables \citep{KMK2013,sanderthreefield,wilson2014fluid,dannberg2016compressible}. 
Regularising the equations in such a way generally means that the system does not exactly reduce to the Stokes problem in the case of zero porosity. Beyond that, \cite{arbogast2017mixed} find that numerical methods that do not take into account the degeneracy of the porosity $\phi$, and instead regularize the equations for example by imposing a small non-zero porosity everywhere, are sure to have a condition number that grows as the porosity approaches zero. 
This makes it computationally expensive to compute numerical models with regions of vanishing porosity. 

\citet{arbogast2017mixed} address this problem by developing a mixed variational framework, carefully scaling the
Darcy variables by powers of the porosity, and defining a mixed finite element method for solving the Darcy--Stokes system. 
This method has the drawback that it requires a particular choice of finite elements, and that it is based on specific assumptions 
on how material properties like the permeability and the bulk viscosity depend on the amount of melt present.
In addition, the method has not been tested on realistic large-scale, 3-D application models or in parallel computations. 

Here, we present a new, more general formulation of the equations for coupled Stokes/Darcy flow that addresses 
these shortcomings, and allows it to compute large-scale 3-D magma/mantle dynamics simulations with extensive regions of zero porosity. 
We have implemented this formulation in the open source geodynamic modelling code \aspect{} \citep{heister2017high,dannberg2016compressible,aspectmanual,aspectsoftware}, which is based on 
the deal.II finite element library \citep{BangerthHartmannKanschat2007, dealII90}. 
Using \aspect{}, we have tested the new method on real-world applications, in parallel, and with adaptive mesh refinement. 

In the following, we will derive our new formulation and its numerical implementation,
and discuss the convergence behaviour that is expected for this method (Section~\ref{sec:formulation}). 
We will demonstrate the correctness of our implementation based on a benchmark case that specifically addresses the boundary between regions with and without melt, and illustrate the improved performance compared to the three-field formulation used in \citet{dannberg2016compressible} (Section~\ref{sec:arbogast_benchmark}). Finally, we will show 2-D and 3-D mid-ocean ridge models to demonstrate the applicability of our method to earth-like settings (Section~\ref{sec:2d-application} and \ref{sec:3d-application}). 
The code used to generate these results can be found in the repository at
\url{https://github.com/geodynamic/aspect} and all input files to reproduce the results are available at \url{https://github.com/tjhei/paper-aspect-melt-paper-2-data}.

\section{Formulation of the problem}
\label{sec:formulation}
We consider the equations describing the behaviour of silicate melt percolating through 
and interacting with a viscously deforming host rock \citep[e.g.][]{McK1984}:

\begin{align}
  \label{eq:McKenzie}
  \frac{\partial}{\partial t} \left[\rho_f \phi \right] + \nabla \cdot \left[ \rho_f \phi \mathbf u_f \right]
  &=
   \Gamma,
  \\  
  \frac{\partial}{\partial t} \left[\rho_s (1 - \phi) \right] + \nabla \cdot \left[ \rho_s (1 - \phi) \mathbf u_s \right]
  &=
  - \Gamma,
  \\[0.5em]
  \label{eq:darcys_law}  
  \phi \left( \mathbf u_f - \mathbf u_s \right)
  &=
  - K_D \left( \nabla p_f - \rho_f \mathbf g \right),
  \\
  \label{eq:stokes-1}
  -\nabla \cdot \left[2\eta \dot{\varepsilon} 
  + \xi (\nabla \cdot \mathbf u_s) \mathbf 1
                \right] + \nabla p_f &=
  \bar{\rho} \mathbf g.
\end{align}

Here, $\phi$ is the porosity, $\rho$ is the density, $\mathbf u$ is the velocity, 
$\Gamma$ is the melting rate, $K_D$ is the Darcy coefficient, $p$ is the pressure, 
$\mathbf g$ is the gravity vector, $\eta$ is the shear viscosity, $\xi$ is the bulk viscosity
and $\dot{\varepsilon} = \nabla \mathbf u_s + (\nabla \mathbf u_s)^T - \frac{1}{3}(\nabla \cdot \mathbf u_s)\mathbf 1$ is the strain rate. 
The index $f$ indicates the melt (fluid), the index $s$ indicates the solid, 
and quantities that are averaged between the solid and the fluid are denoted by a bar. 

Two important material properties in the context of the transition between solid-state mantle convection and 
two-phase magma/mantle dynamics are $K_D$ and $\xi$. 
The Darcy coefficient is defined as the ratio of permeability $k$ and fluid viscosity $\eta_f$, 
and while $\eta_f$ is often, for simplicity, assumed to be constant, the permeability depends on the porosity
as $k\propto\phi^2$ or $k\propto\phi^3$. This means that for vanishing porosity, $K_D=0$. 
The compaction viscosity $\xi$ is often assumed to scale as $\xi \propto \phi^{-1}$, 
so that the matrix can not be compacted ($\xi \rightarrow \infty$) if no melt is present. 

\subsection{Original formulation used in \cite{dannberg2016compressible}}
In previous work \citep{dannberg2016compressible}, we reformulated the equations by building on the three-field formulation from \cite{KMK2013}, extending them to compressible solid and fluid phases:
\begin{align*}
  -\nabla \cdot \left(2\eta \dot{\varepsilon} \right) + \nabla p_f + \nabla p_c  &=
  \bar{\rho} \mathbf g ,
  \\
  \nabla \cdot \mathbf u_s 
  \highlight{
  - \nabla \cdot K_D \nabla p_f 
  - K_D \nabla p_f \cdot \frac{\nabla \rho_f}{\rho_f}
  }
  &= 
  \highlight{
  - \nabla \cdot (K_D\rho_f \mathbf g)  }
  \\
  &\quad\;
  \highlight{+ \Gamma \left( \frac{1}{\rho_f} - \frac{1}{\rho_s} \right)}
  \\
  &\quad\;
  \highlight{- \frac{\phi }{\rho_f} \mathbf u_s \cdot \nabla\rho_f}
  - (\mathbf u_s \cdot \mathbf g ) (1 - \phi) \kappa_s \rho_s
  \\
  &\quad\;
  \highlight{
  - K_D \mathbf g \cdot \nabla \rho_f} ,
  \\
  \nabla \cdot \mathbf u_s + \frac{p_c}{\xi} 
  &=
  0 .
\end{align*}
All terms that vanish in the limit of zero porosity and no melting are highlighted. 
These equations can then be brought into the weak form \citep[see][]{dannberg2016compressible} and solved as outlined in \cite{sanderthreefield}. 
This results in the linear system:
\begin{align} \label{eq:linearsystem}
 \begin{pmatrix}
  \mathbf A & \mathbf B^T & \mathbf B^T \\
  \mathbf B & \mathbf N & \mathbf 0 \\
  \mathbf B & \mathbf 0 & \mathbf K
 \end{pmatrix}
 \begin{pmatrix}
  \mathbf U_s \\ \mathbf P_f \\ {\mathbf P_c}
 \end{pmatrix}
 =
 \begin{pmatrix}
 \mathbf F \\ \mathbf G \\ \mathbf 0  
 \end{pmatrix}.
\end{align}
$\mathbf{A}$, $\mathbf{B}$, $\mathbf{N}$, $\mathbf{K}$, $\mathbf{F}$ and $\mathbf{G}$ are defined as in \cite{dannberg2016compressible}. For a more extensive discussion of these matrix blocks, we refer the reader to Equation~\eqref{eq:linearsystem-new}. 

While this formulation allows it to run large-scale, 3-D models of coupled magma/mantle dynamics, it has several shortcomings:
The number of linear solver iterations increases with an increasing ratio of compaction viscosity $\xi$ and shear viscosity $\eta$, which corresponds to a decreasing porosity $\phi$. 
In addition, for the limit of $\phi \rightarrow 0$ (which implies $\sqrt{K_D} \rightarrow 0$), the compaction pressure $p_c$ is not defined, 
because $\xi \rightarrow \infty$ (the compaction viscosity is generally assumed to scale as $\xi \propto \phi^{-1}$, at least if the model is incompressible, which implies that $-(\mathbf u_s \cdot \mathbf g ) (1 - \phi) \kappa_s \rho_s = 0$). 
In this case, the last two equations become linearly dependent (and the whole system is ill-posed), 
which is also the reason for the increasing number of linear solver iterations that are needed for 
decreasing porosity values. 
Indeed, \cite{arbogast2017mixed} note that all numerical methods that do not specifically take into account the degeneracy of the porosity are sure to have a condition number that grows as the porosity approaches zero. 

In addition, this means that in order to solve the system in spite of this problem, some limit has to be imposed on the compaction viscosity, either in form of a minimum value, or in form of a regularization term that is added to the compaction pressure equation. 
While this form of stabilization allows it to solve the equations, the system will not revert to the incompressible one-phase Stokes equations for vanishing porosity, as there will always be a non-zero contribution of the compaction term that is needed to stabilize the system. 

\subsection{New formulation}
\label{sec:new-formulation}
To address these problems, we have developed a new formulation based on the ideas presented in \citet{arbogast2017mixed}. 
We replace $p_c$ by $\bar{p_c}$, using the relation $p_c = \sqrt{K_D'} \bar{p_c}$, where $K_D'$ is the Darcy coefficient, 
but rescaled to a reference value representative of the model (for details, see Section~\ref{sec:constraining-K-D}). 
To keep the matrix symmetric, we also scale the last equation by $\sqrt{K_D'}$, 
and arrive at the following, new system of partial differential equations:
\begin{align*}
  -\nabla \cdot \left(2\eta \dot{\varepsilon} \right) 
  + \nabla p_f 
  \highlight{+ \nabla \left(\sqrt{K_D'} \bar{p_c}\right)}
  &=
  \bar{\rho} \mathbf g ,
  \\
  \nabla \cdot \mathbf u_s 
  \highlight{
  - \nabla \cdot K_D \nabla p_f 
  - K_D \nabla p_f \cdot \frac{\nabla \rho_f}{\rho_f}
  }
  &= 
  \highlight{- \nabla \cdot (K_D\rho_f \mathbf g)}
  \\
  &\quad\;
  \highlight{+ \Gamma \left( \frac{1}{\rho_f} - \frac{1}{\rho_s} \right)}
  \\
  &\quad\;
  \highlight{- \frac{\phi }{\rho_f} \mathbf u_s \cdot \nabla\rho_f}
  - (\mathbf u_s \cdot \mathbf g ) (1 - \phi) \kappa_s \rho_s
  \\
  &\quad\;
  \highlight{- K_D \mathbf g \cdot \nabla \rho_f} ,
  \\
  \highlight{\sqrt{K_D'} \nabla \cdot \mathbf u_s + \frac{K_D' \bar{p_c}}{\xi}}
  &=
  0 .
\end{align*}
Again, terms that vanish in the limit of zero porosity are marked in gray boxes. 
For this new formulation, it becomes apparent that for the limit of $\phi \rightarrow 0$, the last equation vanishes completely and we recover the Stokes system from the first two equations, as $\nabla (\sqrt{K_D'} {\bar p}_c) \rightarrow 0$. 
\begin{align*}
  -\nabla \cdot \left(2\eta \dot{\varepsilon} \right) 
  + \nabla p_f &=
  \bar{\rho} \mathbf g ,
  \\
  \nabla \cdot \mathbf u_s 
  &=
  - (\mathbf u_s \cdot \mathbf g ) \kappa_s \rho_s.
\end{align*}
In contrast to the original formulation, the real compaction pressure $p_c$ is now defined everywhere, 
as it is computed as $p_c = \sqrt{K_D'} \bar{p_c}$. So for the case of vanishing melt fraction $\phi = 0$, 
which implies $\sqrt{K_D'}=0$, this scaling always leads to $p_c=0$. 
While the (rescaled) compaction pressure $\bar{p_c}$ is still not defined in this limit, 
it is also not used anywhere in the system. Hence, to make sure that the linear system can be solved, 
we can constrain these degrees of freedom in regions where the porosity is below a given threshold to $\bar{p_c}=0$. 
An example for this is given in Figure~\ref{fig:melt_cells}. 
Beyond that, this formulation has the advantage that no additional computational resources are used 
to solve the coupled Stokes/Darcy system if no melt is present. 

Solving the single phase Stokes system if the porosity is below a given threshold can also be motivated physically: 
When solid rock starts to melt, melt is expected to form in isolated patches between the mineral grains. 
Melt segregation and compaction only start to occur once the porosity reaches a critical value -- the percolation threshold -- and pockets of melt become interconnected \citep[e.g.][]{zhu2003network, cheadle2004percolation}. In other words, the permeability of the rock is zero, and there is no relative movement between the phases, until a given porosity is reached. 
Below that percolation threshold, the motion of the rock can be described accurately by single phase Stokes flow. 
As the critical porosity is influenced by grain size, composition, and other properties of the rock, 
it can be chosen as a model input parameter. 

In the incompressible formulation, which is a good approximation for models that do not span a large depth range and is commonly used for these applications, all terms that contain the solid compressibility or the fluid density gradient vanish: 
\begin{align}
  \label{eq:new-form-incompressible-1}
  -\nabla \cdot \left(2\eta \dot{\varepsilon} \right) 
  + \nabla p_f + \nabla (\sqrt{K_D'} \bar{p_c})  
  &=
  \bar{\rho} \mathbf g ,
  \\
  \label{eq:new-form-incompressible-2}
  \nabla \cdot \mathbf u_s - \nabla \cdot K_D \nabla p_f 
  &= 
  - \nabla \cdot (K_D\rho_f \mathbf g)
  \notag
  \\
  &\quad
  + \Gamma \left( \frac{1}{\rho_f} - \frac{1}{\rho_s} \right),
  \\
  \label{eq:new-form-incompressible-3}
  \sqrt{K_D'} \nabla \cdot \mathbf u_s + \frac{K_D' \bar{p_c}}{\xi} 
  &=
  0 .
\end{align}

The weak form of the full problem is given by: Find $\mathbf u_s, p_f, p_c$ with
\begin{align}\label{eq:finalweakform1}
 \left(2\eta \dot{\varepsilon}(\mathbf u_s), \dot{\varepsilon}(\mathbf v_s) \right)
  -  \left( \frac{2}{3}\eta \nabla \cdot \mathbf u_s, \nabla \cdot \mathbf v_s \right) &
  \notag \\
  - (p_f, \nabla  \cdot \mathbf v_s)
  - (\sqrt{K_D'} \bar{p_c}, \nabla \cdot \mathbf v_s) 
  &=
  \left(\bar{\rho} \mathbf g, \mathbf v_s \right) ,
  \\
  \label{eq:finalweakform2}
  - \left( \nabla \cdot \mathbf u_s, q_f \right) 
  - \left(K_D \nabla p_f , \nabla q_f \right) \qquad &  \notag \\
  + \left( K_D \nabla p_f \cdot \frac{\nabla \rho_f}{\rho_f} , q_f \right)
  &= 
  - \left( K_D\rho_f \mathbf g, \nabla q_f \right)
  \notag \\
  &\quad
  + \int_{\partial \Omega} q_f K_D (\rho_f \mathbf g - \mathbf f_2) \cdot \vec{n} \;\text{d}s
  \notag \\
  &\quad
  - \left( \frac{1}{\rho_f} - \frac{1}{\rho_s} \right)
  (\Gamma, q_f )
  \\
  &\quad
  + \left( \frac{\phi }{\rho_f} \mathbf u_s \cdot \nabla\rho_f , q_f \right)
  \notag \\
  &\quad
  + \left( (\mathbf u_s \cdot \mathbf g ) (1 - \phi) \kappa_s \rho_s , q_f \right)
  \notag \\
  &\quad
   + \left( K_D \mathbf g \cdot \nabla \rho_f, q_f \right) ,
  \notag \\
 - \left( \sqrt{K_D'} \nabla \cdot \mathbf u_s, q_c \right)
  -  \left( \frac{1}{\xi} K_D' \bar{p_c}, q_c \right)
  &=
  0 . \label{eq:finalweakform3}
\end{align}
for all test functions $\mathbf v_s, q_f, q_c$.

Note that we have made the assumption that at the interface $\partial \Omega_\text{melt}$ between regions 
where the compaction pressure is constrained to $p_c=0$ and the regions where we solve for the full two-phase system,  
$\nabla p_f = \rho_f \mathbf g$. 
This follows from integration by parts of Equation~\eqref{eq:new-form-incompressible-2}, which yields
\begin{align*}
  - \left( \nabla \cdot \mathbf u_s, q_f \right) 
  - \left(K_D \nabla p_f , \nabla q_f \right)
  &= 
  - \left( K_D\rho_f \mathbf g, \nabla q_f \right)
  - \left( \frac{1}{\rho_f} - \frac{1}{\rho_s} \right) (\Gamma, q_f ) \\
  &\quad
  + \int_{\partial \Omega_\text{melt}} q_f K_D (\rho_f \mathbf g - \nabla p_f) \cdot \vec{n} \;\text{d}s
\end{align*}
for the interface $\partial \Omega_\text{melt}$. 
As $K_D=0$ in the cells without melt, and $K_D>0$ in the cells where melt is present,
$\int_{\partial \Omega_\text{melt}} (\rho_f \mathbf g - \nabla p_f) \cdot \vec{n} \;\text{d}s = 0$.
Because of Darcy's law (Equation~\ref{eq:darcys_law}), this condition is equivalent to the assumption that the melt velocity equals 
the solid velocity at the interface between the two regions. 

This means we have to solve the linear system:
\begin{align} \label{eq:linearsystem-new}
 \begin{pmatrix}
  \mathbf A & \mathbf B^T & \sqrt{K_D'}\mathbf B^T \\
  \mathbf B & \mathbf N & \mathbf 0 \\
  \sqrt{K_D'}\mathbf B & \mathbf 0 & K_D'\mathbf K
 \end{pmatrix}
 \begin{pmatrix}
  \mathbf U_s \\ \mathbf P_f \\ \bar{\mathbf P_c}
 \end{pmatrix}
 =
 \begin{pmatrix}
 \mathbf F \\ \mathbf G \\ \mathbf 0  
 \end{pmatrix},
\end{align}
where $\mathbf A$ is the discretization of $\left(2\eta \dot{\varepsilon}(\mathbf u_s), \dot{\varepsilon}(\mathbf v_s) \right)
  -  \left( \frac{2}{3}\eta \nabla \cdot \mathbf u_s, \nabla \cdot \mathbf v_s \right)$, 
$\mathbf B$ is given by $- (p_f, \nabla  \cdot \mathbf v_s)$,
$\mathbf F$ is given by $\left(\bar{\rho} \mathbf g, \mathbf v_s \right)$,
$\mathbf N$ is given by $-\left(K_D \nabla p_f , \nabla q_f \right)$ in the incompressible case, 
$\mathbf G$ is given by $- \left( K_D\rho_f \mathbf g, \nabla q_f \right) + \int_{\partial \Omega} q_f K_D (\rho_f \mathbf g - \mathbf f_2) \cdot \vec{n} \;\text{d}s - \left( \frac{1}{\rho_f} - \frac{1}{\rho_s} \right) (\Gamma, q_f )$ in the incompressible case, 
and $\mathbf K$ is given by $-\left( \frac{1}{\xi} \bar{p_c}, q_c \right)$. For compressible computations, $\mathbf N$ also contains the
non-symmetric, third term from \eqref{eq:finalweakform2}, and $\mathbf G$ contains the remaining terms on the right-hand side of \eqref{eq:finalweakform2}, which contain $\kappa_s$ and $\nabla \rho_f$.

As the block structure of the linear system remains the same as in \cite{dannberg2016compressible}, 
the same solver strategy, based on \cite{sanderthreefield}, can be employed to solve the block system \eqref{eq:linearsystem-new}. 
Specifically, we use flexible GMRES with the block preconditioner (preconditioned from the right):
\begin{align*}          
 \mathbf P^{-1} = \begin{pmatrix}
 \mathbf A & \mathbf B^T & \sqrt{K_D'} \mathbf  B^T \\
 \mathbf 0 & \mathbf -\frac{1}{\eta} \mathbf M_{p_f} - K_D \mathbf
 L_{p_f}   & -\frac{\sqrt{K_D'}}{\eta} \mathbf M_{p_f,p_c} \\
 \mathbf 0 & \mathbf -\frac{\sqrt{K_D'}}{\eta} \mathbf M_{p_c,p_f}   &
 \mathbf -K_D' (\frac{1}{\eta} + \frac{1}{\xi}) \mathbf M_{p_c}
 \end{pmatrix} ^{-1}.
 \end{align*}
$\mathbf M_*$ and $\mathbf L_*$ are mass and stiffness matrices, respectively.
$\mathbf A^{-1}$ is approximated using an inner CG solver preconditioned 
by Trilinos ML applied to the diagonal blocks of $\mathbf A$. The Schur complement solves for 
the bottom-right 2 by 2 block are also done using CG preconditioned by Trilinos ML.

\subsection{Constraining the compaction pressure DoFs}
\label{sec:constraining-K-D}

As outlined in Section~\ref{sec:new-formulation}, we constrain the compaction pressure degrees of freedom to $\bar{p_c}=0$ in regions where the porosity is below a given threshold.
In practice, we choose this threshold $K_\text{threshold}$ based on the Darcy coefficient $K_D$ relative to a reference value $K_{D_0}$, 
as this ratio is what we use to rescale the different matrix blocks in the linear system~\eqref{eq:linearsystem-new}. 
$K_{D_0}$ is defined as the ratio of permeability and fluid viscosity at a porosity that is typical for the model 
(in the following examples, we will use a value of 1\%), but because it is part of \aspect{}'s `material model' plugin structure
\citep[see][]{aspectmanual} it can be chosen in dependence of the specific application case. 
This means that the last equation in \eqref{eq:linearsystem-new} will not be rescaled at all if the porosity equals this reference porosity.
 
\begin{figure}
 \centering
 \includegraphics[width=0.8\textwidth]{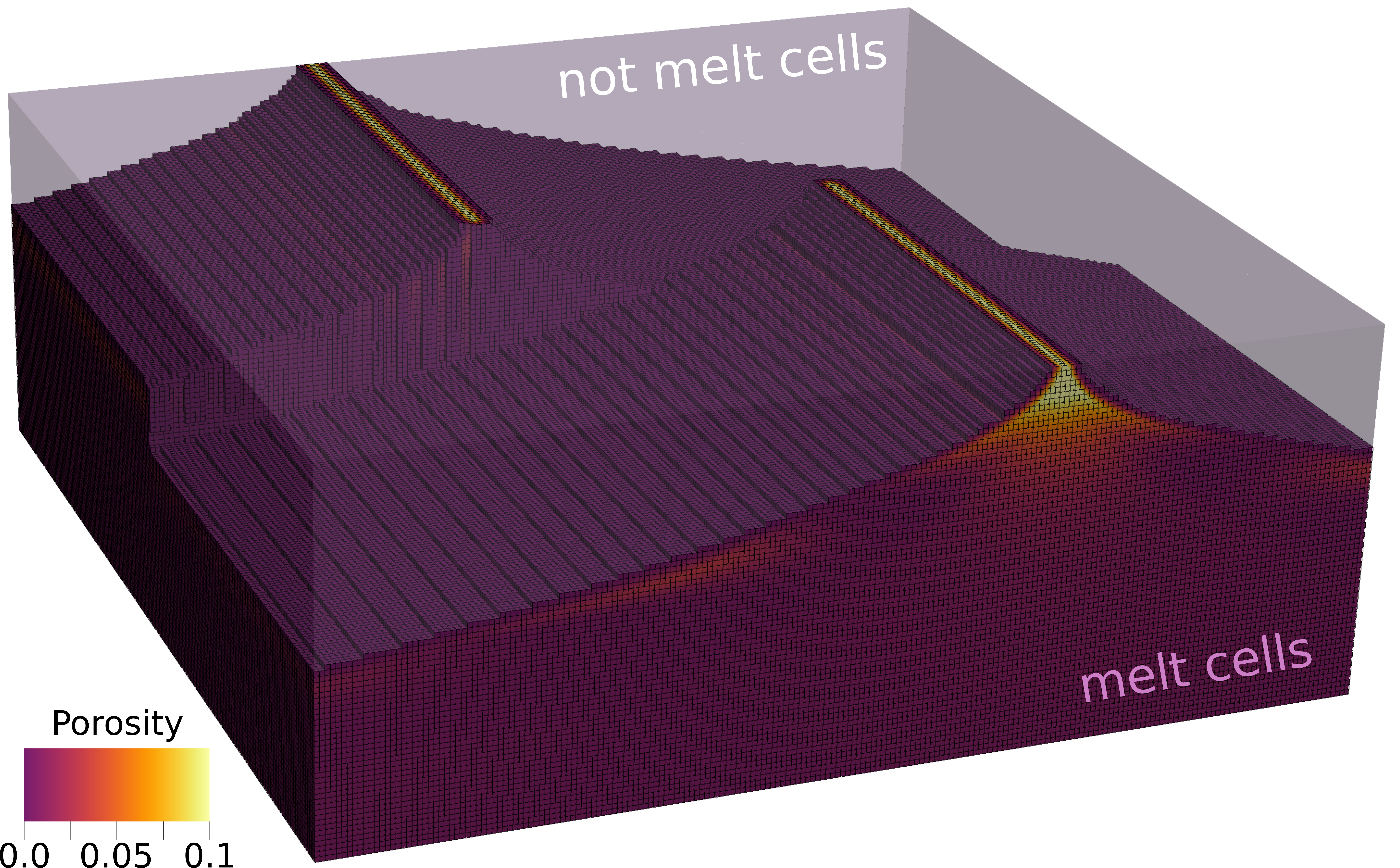}
 \caption{Distribution of melt cells an cells that are not melt cells in a 3-D model of a transform fault. 
          The coupled Stokes--Darcy equations are only solved in cells where the porosity is above a given threshold.}
  \label{fig:melt_cells}
\end{figure}

The decision to constrain degrees of freedom is made for each cell, separating the model domain into `melt cells', where the full equations are solved,
and cells that are not `melt cells' with the compaction pressure degrees of freedom being constrained. 
An example for this is shown in Figure~\ref{fig:melt_cells}. 
A cell is determined to be a `melt cell' if $K_D / K_{D_0} > K_\text{threshold}$ on any of the quadrature points. 
The default value is given by $K_\text{threshold} = 10^{-3}$, but it is an input parameter that can be chosen differently in each model 
(for its influence on solver performance, see Section~\ref{sec:material-properties}). 
Based on this evaluation, a scaling factor $\sqrt{K_D'}$ for the compaction pressure is computed for each cell. 
In melt cells, $K_D' = \max{(K_{D_\text{mean}} / K_{D_0}, K_\text{threshold})}$, where $K_{D_\text{mean}}$ is the geometric mean 
of the Darcy coefficient for the respective cell. 
Providing a minimum value for the scaling factor guarantees that we avoid the mathematically degenerate region in all quadrature points where we solve the two-phase flow equations.
In cells that are not melt cells, we set $K_D'=0$, and all compaction pressure degrees of freedom are constrained to zero. Effectively, this removes the equations for $\bar{p_c}$ in the Stokes region
and for a computation without any melt cells, the linear system and solver cost is effectively
equivalent to a standard Stokes solver.

This algorithm is executed once in every time step, after solving the advection equation for the porosity, to make sure that the constraints for system~\eqref{eq:linearsystem-new} are the same for every nonlinear iteration and that the nonlinear solver converges. 
To compute the Darcy coefficient in Equation~\eqref{eq:finalweakform2}, the same threshold is applied:
$K_D = \max{(K_{D_\text{mean}}, K_\text{threshold} \, K_{D_0})}$ in melt cells, and zero otherwise. 

\cite{dannberg2016compressible} used a different threshold to discriminate between model regions with and without melt migration, 
directly based on the porosity. In their method, the full two-phase flow equations are only solved for $\phi > \phi_\text{threshold}$. 
Both methods are compared for different threshold values in Section~\ref{sec:material-properties}, and -- assuming a reference porosity of 0.01 -- both thresholds are related as $\phi_\text{threshold} = 0.01 {K_\text{threshold}}^{2/3}$.

\subsection{Finite Element Formulation}
\label{sec:analysis}

Now we still require a choice of finite element spaces for the discrete solution
$(\mathbf u_{s,h},p_{f,h},p_{c,h}) \in W_h \subset W = H_0^1(\Omega) \times H^1(\Omega) \times L^2(\Omega)$.
This system is analysed in detail in \citep{grove2017discretizations}.
We use quadrilateral cells and the following, typical polynomial finite element spaces:
Let $Q_k$ be the continuous space with tensor-product polynomials of degree $k$ on each cell
and let $DGP_k$ be the discontinuous space with polynomials of degree $k$.

We choose $Q_{k+1}$ for each component of the velocity $u_{s,h}$. To be able to
solve for a discrete $p_{c,h}$ in \eqref{eq:finalweakform3}, the space needs to
to be discontinuous to allow
a jump from melt to a no-melt cell, so we choose $DGP_k$.
For $p_{f,h}$ we have two sensible choices:
\[
 W_k = Q^d_{k+1} \times Q_{k+1} \times DGP_k
 \qquad \text{or} \qquad 
% \]
% or
% \[
 W^L_k = Q^d_{k+1} \times Q_{k} \times DGP_k,
\]
namely choosing a higher or lower polynomial degree.

If we consider the case $K_D=0$ (no melt in the domain), we recover the standard Stokes system and
well-posedness requires a stable finite element choice for $\mathbf u_{s,h}$ and $p_{f,h}$ to
guarantee convergence. One example is the usage of Taylor-Hood elements,
where the velocity is discretized with one polynomial degree higher than the pressure
like in the definition of $W^L_k$. This means $W^L_k$ is stable and gives optimal convergence rates,
while $W_k$ is not a stable choice leading to oscillating solutions.

On the other hand, if we consider a situation with melt everywhere ($K_D \geq K_{D,\text{min}} > 0$),
both discrete spaces give stable solutions, but $W^L_k$ gives suboptimal convergence
rates, while $W_k$ will achieve optimal rates provided the data and exact solution
are sufficiently smooth.
We note that the inverse of the minimum value of $K_D$ appears in the stability estimate,
confirming the issue of letting $K_D$ go to zero.

\begin{figure}
 \centering
 \includegraphics[width=0.8\textwidth]{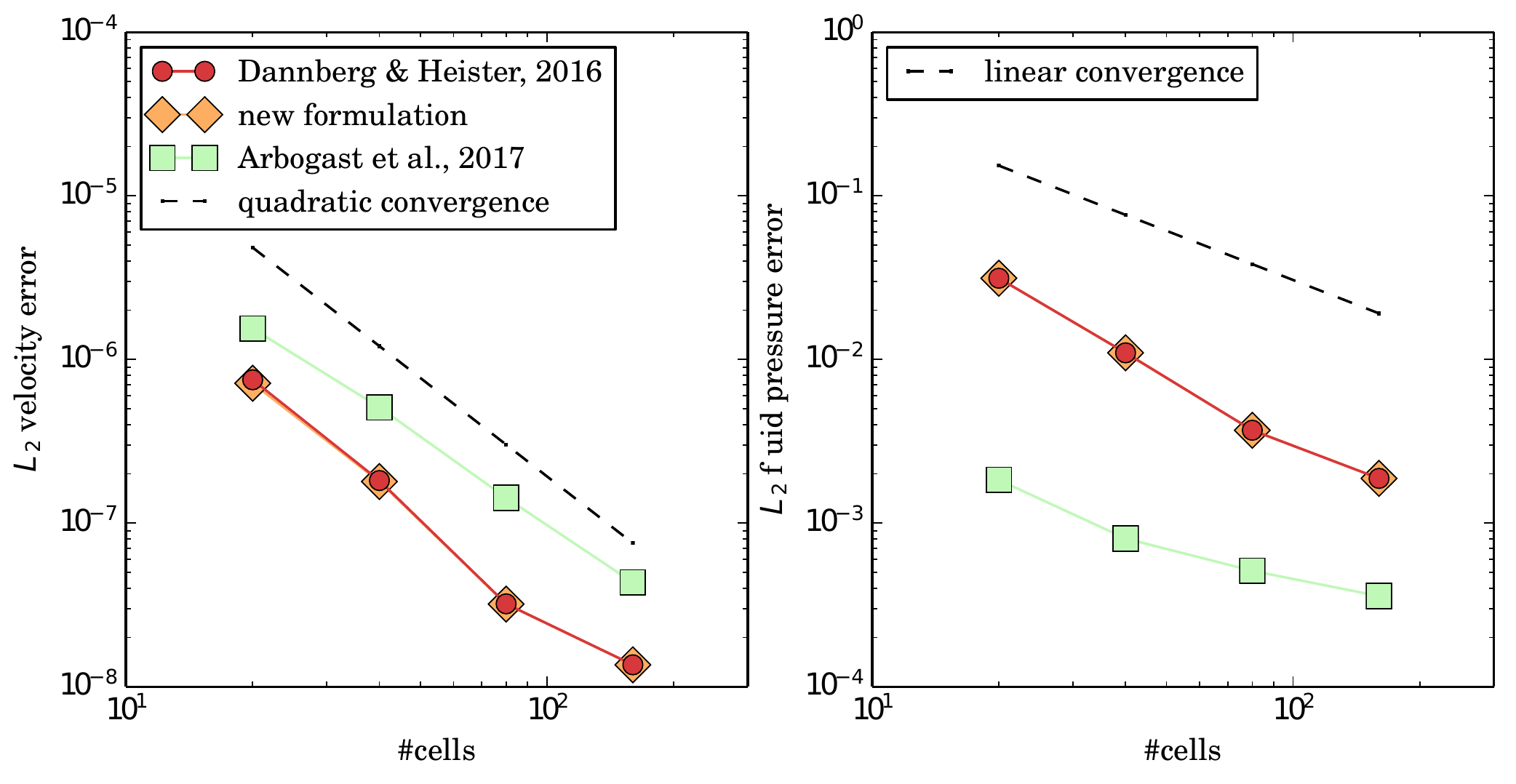}
 \caption{Error for solid velocity and fluid pressure. Results from \protect\cite{arbogast2017mixed}, Table 6 are plotted for comparison. For the method of \protect\cite{dannberg2016compressible}, the linear solver does not converge for a resolution higher than $n=80$, so the results shown are using a direct solver.}
  \label{fig:1D_analytical_solution}
\end{figure}

We decided to do our computations with $W^L_2$, the lowest order discretization
that gives stable solutions even for $K_D=0$. Table~\ref{table:fem} summarizes 
the discretization choices and convergence rates in $L_2(\Omega)$, and
the convergence rates in Figure~\ref{fig:1D_analytical_solution} confirm these
estimates.

Alternatives would be to either always require a minimum $K_D$, add stabilization
terms to make the Stokes solution stable for $K_D=0$, or discretize with different
finite element spaces in the regions with and without melt.

\begin{table}
\begin{center}
\begin{tabular}{l|lll}
definition & optimal rates for $\mathbf u_s,p_f,p_c$ & $K_D=0$ & $K_D>0$ \\ \hline
$W_2 = Q^d_{2} \times Q_{2} \times DGP_1$ & 3,3,2 & unstable & 3,3,2 (optimal)\\
$W^L_2 = Q^d_{2} \times Q_{1} \times DGP_1$ & 3,2,2 & 3,2,- (optimal) & 2,2,1 (suboptimal)
\end{tabular}
\end{center}
  \caption{Different choices for finite element spaces. The columns contain the 
  optimal convergence rates based on approximation quality of the element in the 
  L2 norm, the expected convergence rates for a problem without melt, and
  for melt everywhere.}
\label{table:fem} 
\end{table}

\section{Results}

\subsection{1-D analytical solution for the interface between regions with and without melt}
\label{sec:arbogast_benchmark}

\begin{figure}
 \centering
 \includegraphics[width=\textwidth]{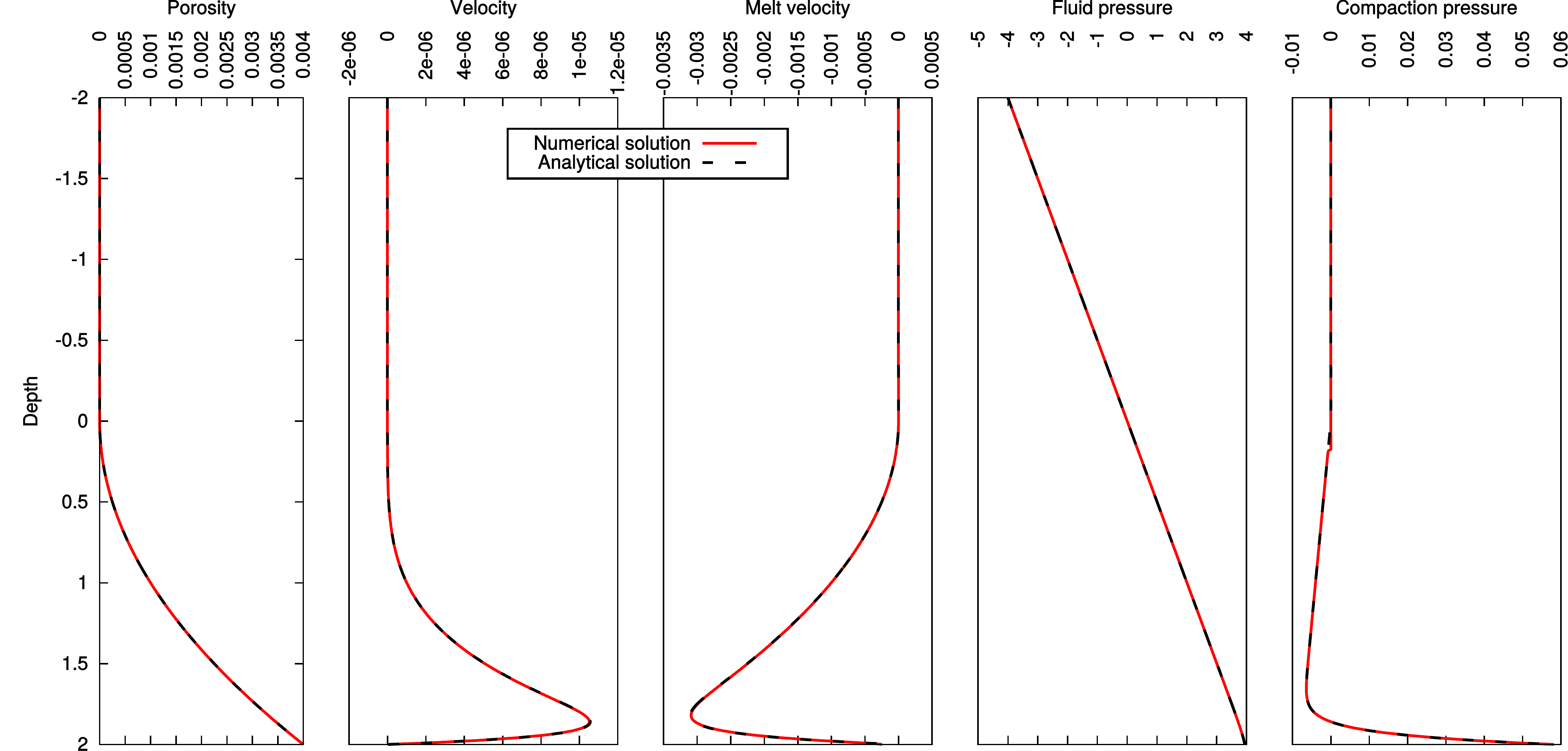}
 \caption{Setup of the benchmark given in \cite{arbogast2017mixed}. The solution derived in \cite{arbogast2017mixed} is given as a dashed black line, and the solution computed numerically with \aspect{} is marked by a red line.}
  \label{fig:1D_setup}
\end{figure}

We use a 1-D benchmark from \cite{arbogast2017mixed} to show that our formulation is correct, 
and that the solver performs much better than the previous one in \cite{dannberg2016compressible}. 
The benchmark specifically addresses the transition between regions with both melt and solid
-- where the coupled Stokes/Darcy systems is solved -- 
and regions without melt -- where the problem is reduced to the Stokes problem. 
This is done by choosing the porosity as zero in the upper half of the model domain, and as a quadratic function 
in the lower half, in such a way that the transition between the two regions is continuous and smooth (Figure~\ref{fig:1D_setup}). 
Under the assumption that $\phi \ll 1$, \citet[Equations 6.21--6.23]{arbogast2017mixed} derive an approximate solution for this given porosity 
distribution, which we use to compute errors and convergence rates of our method. 

Our numerical results show similar convergence rates as \cite{arbogast2017mixed}: quadratic convergence for the solid velocity and linear convergence for the fluid and compaction pressure. 
Beyond that, we find that the number of linear solver iterations is not sensitive to problem size, 
and that the iteration count does not vary substantially in dependence of the material properties, 
such as, for example, the ratio between shear and compaction viscosity (Tables~\ref{table:1D-iteration-counts} and \ref{table:1D-iteration-counts-parameters}). 
This is a substantial improvement from the very strong dependence on both problem size and material properties exhibited by the method used in \cite{dannberg2016compressible}, which is what motivated the present study.

\begin{table}
  \centering
  \begin{tabular}{|c|rr|}
    \multicolumn{3}{c}{\textbf{Problem size: Number of linear solver iterations}}  \\
    \hline
    \#cells & Dannberg \& Heister (2016) & this study \\ \hline
    $20$ & 107 & 5 \\
    $40$ & 303 & 7 \\
    $80$ & 820 & 10 \\
    $160$ & no convergence & 8 \\
    \hline
  \end{tabular}
  \caption{Iteration count in dependence of the problem size. While for the method of \cite{dannberg2016compressible} the number of iterations increases with the number of degrees of freedom, our new method needs fewer iterations and the iteration count is independent of the problem size.}
\label{table:1D-iteration-counts}
\end{table}

\begin{table}
  \centering
  \begin{tabular}{|c|rr|}
	\multicolumn{3}{c}{\textbf{Parameters variations: Number of linear solver iterations}}  \\
	\hline
	$\xi_\text{max}/\eta$ & Dannberg \& Heister (2016) & this study \\ \hline
	$10^1$ & 24 & 11 \\
	$10^2$ & 63 & 12 \\
	$10^3$ & 214 & 14 \\
	$10^4$ & 820 & 16 \\
	$10^5$ & no convergence & 16 \\
	$10^6$ & no convergence & 16 \\
	$10^7$ & no convergence & 11 \\
	\hline
  \end{tabular}
  \caption{Iteration count in dependence of the bulk-to-shear-viscosity ratio, for $n=80$ cells in vertical direction.}
  \label{table:1D-iteration-counts-parameters}
\end{table} 

\subsection{Numerical results: 2-D mid-ocean ridge model}
\label{sec:2d-application}

\begin{figure}
 \centering
 \includegraphics[width=\textwidth]{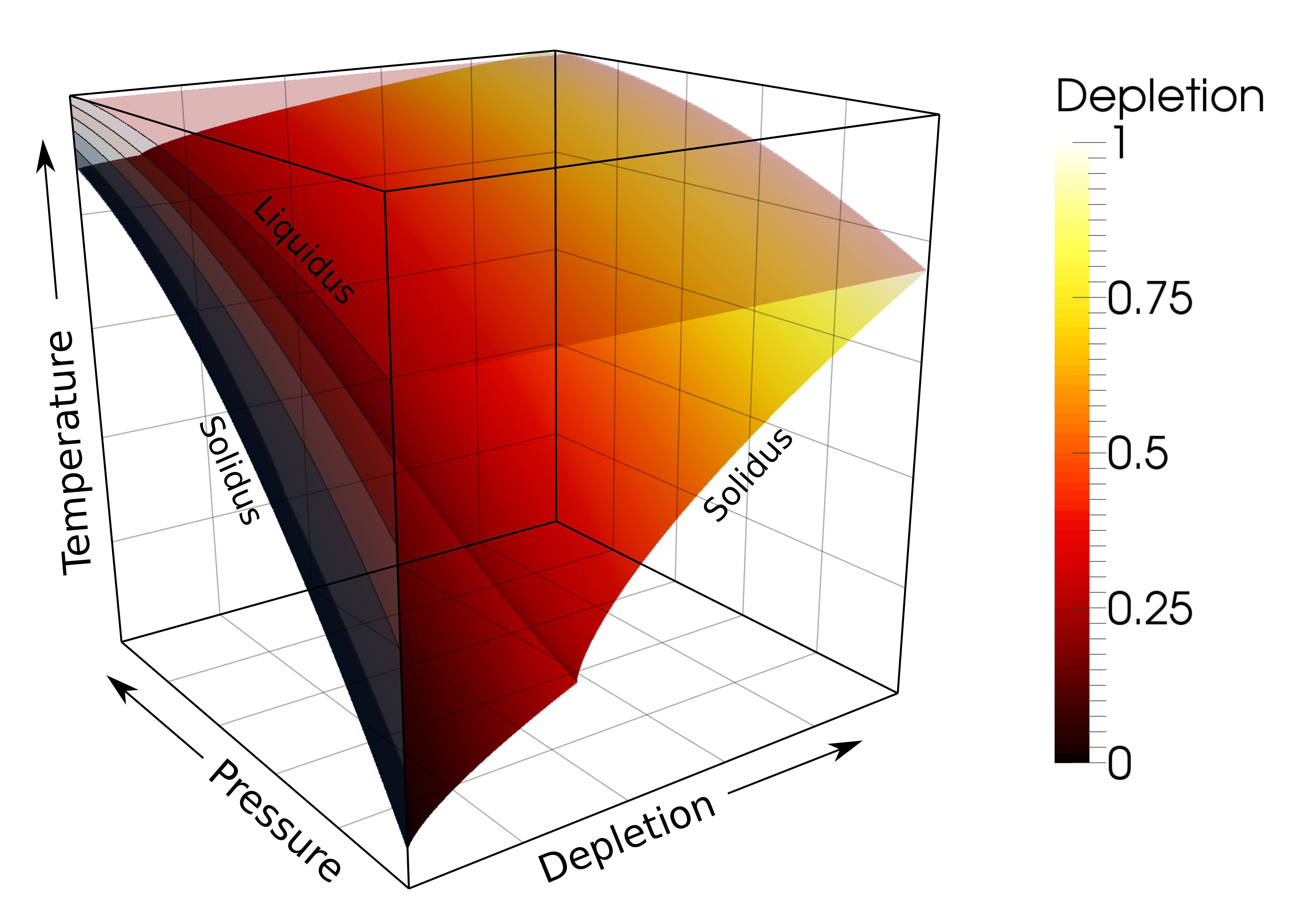}
 \caption{Melting parametrization from \cite{KSL2003}. Shown is a temperature range from 1300 to 2300 K and a pressure range from 0 to 10 GPa. The kink signifies the exhaustion of clinopyroxene in the host rock. 
 Contours between solidus and liquidus in the temperature--pressure plane are drawn at melt fractions of 0\%, 20\%, 40\%, 60\%, 80\%, and 100\%.}
  \label{fig:melting_parametrization}
\end{figure}

In the previous section we have shown that our formulation correctly reproduces analytical solutions 
and solver performance is independent of problem size and contrast between shear and compaction viscosity. 
In the following, we will demonstrate that our implementation also performs well for realistic models of coupled magma/mantle dynamics that are relevant for advancing our understanding of how magma rises from its source region to the surface. 
For this purpose, we set up a mid-ocean ridge model with a visco-plastic, temperature and porosity-dependent rheology. 
Prescribed outflow at the side boundaries leads to corner flow within the domain, so that inflowing material rises and melts adiabatically below the ridge. 
We use the melting parametrization from \cite{KSL2003} as depicted in Figure~\ref{fig:melting_parametrization}. 
To track the temperature, the porosity and the degree of melting (depletion), we use second-order finite elements 
and advect them as fields, employing an entropy-viscosity stabilization \citep{GPP11}. 

\subsubsection{Boundary conditions}

\begin{figure}
 \centering
 \includegraphics[width=0.5\textwidth]{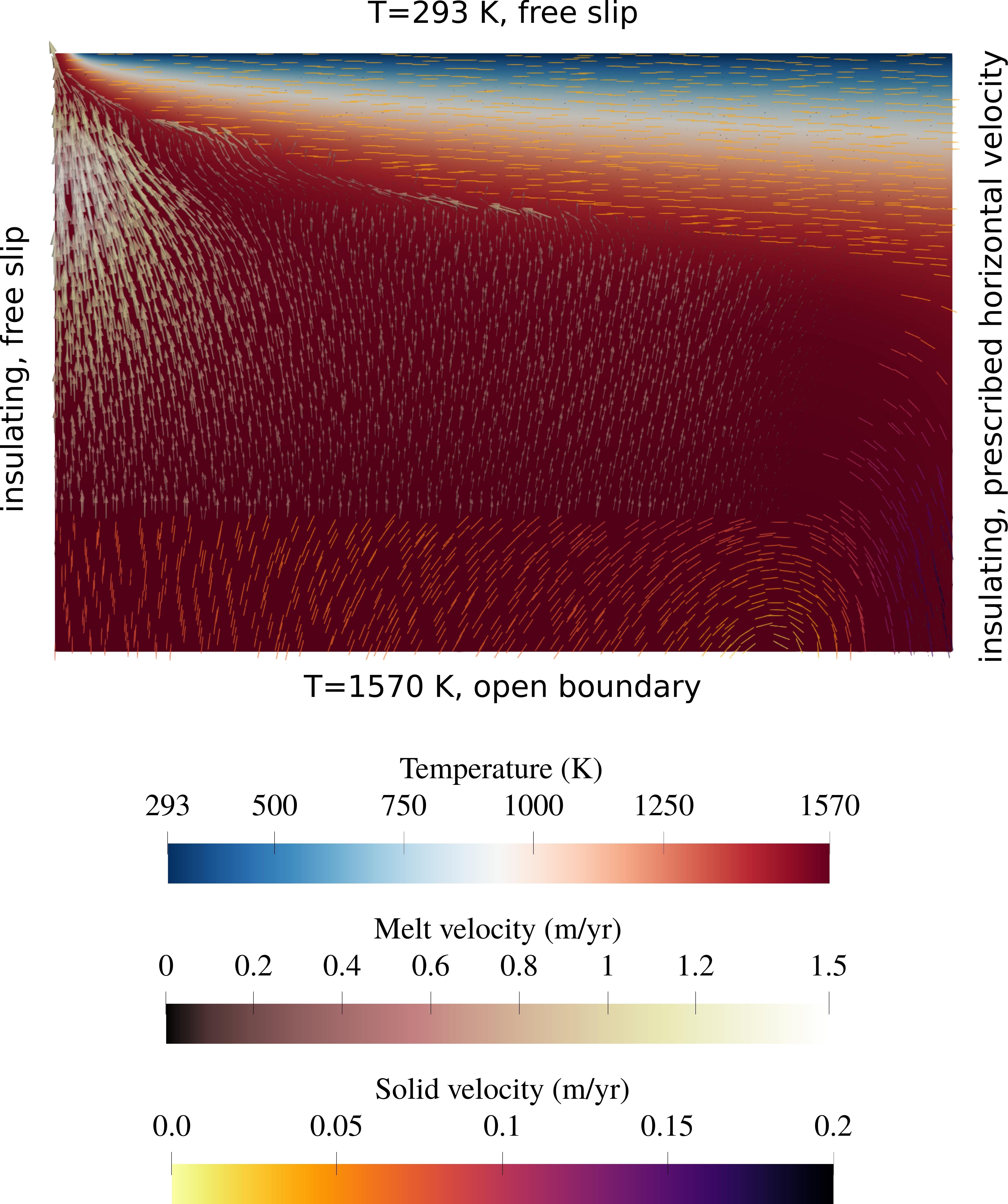}
 \caption{Setup of the mid-ocean ridge model. Arrows mark the melt velocity and lines illustrate the direction of the solid velocity.}
  \label{fig:ridge-setup}
\end{figure}

The temperature is fixed to 293~K at the top boundary and to 1570~K at the bottom boundary, while the side boundaries are insulating. Porosity and depletion fields are fixed to zero at the inflow (bottom) boundary, and Neumann boundary conditions are applied at the other boundaries. 
We prescribe the horizontal component of the velocity to a constant value of 4 cm/yr on the right model boundary to generate the corner flow that is typical for mid-ocean ridges. In addition, the lithostatic pressure is applied as a traction boundary condition for the vertical stress component at the right boundary and the stress at the bottom boundary, allowing free in and outflow. The top and left boundaries are free-slip boundaries and are impermeable to flow. Figure~\ref{fig:ridge-setup} illustrates the setup. 

To allow melt to escape at the ridge axis, we add a temperature perturbation to the otherwise constant boundary temperature at the top of the model in form of a hyperbolic tangent close to the ridge axis.
So the total boundary temperature is defined as 
\begin{equation}
T = T_0 + \Delta T \left(1 - \tanh\left(\frac{x-x_0}{w}\right)\right),
\end{equation}
with $T_0 = 293$\,K, $\Delta T = 600$\,K, $x_0 = 2000$\,m and $w = 1000$\,m. 
This leads to a nonzero melt fraction at the ridge axis, where melt can flow out of the model domain. 
To avoid a suction effect at the ridge axis, 
we prescribe the fluid pressure gradient at the upper model boundary as 
\begin{equation}
\nabla p_f = \left(f \rho_f + (1-f) \rho_s\right) \mathbf g,
\end{equation}
similarly to the pressure boundary condition applied in \cite{Katz2010} for the same reason. 
In this expression, $f$ controls the resistance to flow through the boundary: 
$f=0$ would allow free outflow of melt like at an open boundary, and $f=1$ corresponds to a closed boundary. 
Here we use $f=0.99$, which is large enough to let melt flow out of the domain and limit the steady-state porosity beneath the ridge axis to approximately 10\%, but not so large that the outflow dominates the melt flow in the whole melting region. 

\subsubsection{Material properties}
\label{sec:rheology}

We combine a temperature and porosity-dependent diffusion creep rheology with a stress limiter of the following form:
\begin{align}
\eta(\phi,T) = 
\begin{cases}
\eta_0 e^{-\alpha\phi -\beta(T-T_0)/T_0},& \text{if } \sigma < \sigma_{yield}\\
\frac{\sigma_{yield}}{2 \dot\varepsilon_{II}} & \text{otherwise}
\end{cases}
\end{align}
where $\dot\varepsilon_{II}$ is the second invariant of the strain rate, and $\sigma_{yield} = C \cos(\phi_{\eta}) + p_s \sin(\phi_{\eta})$ with the cohesion $C$ and the friction angle $\phi_{\eta}$.
This way, the stress will not exceed the yield strength of the material, and deformation is localized 
at the ridge axis. 
The compaction viscosity is given as
\begin{align}
\xi(\phi,T) = \xi_0 \frac{\phi_0}{\phi} \, e^{-\beta(T-T_0)/T_0}, 
\end{align}
with the reference porosity $\phi_0=$~0.05.

Most other material properties are chosen as in the mid-ocean ridge model in \cite{Katz2010}. 
The model is incompressible, so that the density is given as
\begin{align}
\rho = \left[ (\rho_s + \Delta\rho_C F) (1 - \phi) + \rho_f \phi \right] (1 - \alpha_\text{thermal} (T - T_\text{ref})),
\end{align}
where $F$ is the degree of melting (depletion), $\Delta\rho_C$ is the density change due to depletion, 
and $\alpha_\text{thermal}$ is the thermal expansivity. 
A complete list of input parameters is given in Table~\ref{table:parameters}. 

\begin{table*}
\centering
\caption{Parameters used for the mid-ocean ridge models.}
\begin{tabular}{ ll} 
 \hline
 \hline
 Quantity     & Value\\ 
 \hline
 Reference bulk viscosity  $\xi_0$      & $4 \cdot 10^{20}$ Pa s\\
 Reference shear viscosity $\eta_0$     & $10^{18}$ Pa s\\
 Melt viscosity $\eta_f$                & $1$ Pa s\\
 Solid density $\rho_{s}$               & $3000~\text{kg/m}^3$\\
 Fluid density $\rho_{f}$               & $2500~\text{kg/m}^3$\\
 Compositional density contrast $\Delta\rho_C$ & $500~\text{kg/m}^3$\\
 Reference permeability $k_0$           & $10^{-7} \text{ m}^2$\\
 Reference porosity $\phi_0$            & $0.05$ \\
 Melt weakening parameter $\alpha$      & $27$\\
 Temperature weakening parameter $\beta$  & $24$\\
 Thermal expansivity $\alpha_\text{thermal}$ & $2 \times 10^{-5}$ 1/K\\
 Specific heat $C_p$ & $1250$ J/(kg K)\\
 Reference temperature $T_\text{ref}$   & 1600 K\\
 Thermal conductivity $k_\text{thermal}$ & 4.7 W/(m\,K)\\
 Cohesion $C$                           & $2 \cdot 10^{7}$ Pa\\
 Friction angle $\phi_{\eta}$           & 30\degree\\
 \hline
 X extent     & 105 km\\
 Z extent     & 70 km\\
 \hline
 \hline
\end{tabular}
\label{table:parameters}
\end{table*}

\subsubsection{Initial conditions}

We first run a time-dependent model to generate realistic temperature, composition and porosity distributions for our scaling tests, which are instantaneous.

To prescribe initial conditions for the temperature and composition in the time-dependent model, we use a temperature distribution based on the half-space cooling model to compute the equilibrium melt fraction everywhere in the domain. As we take into account latent heat effects, this initial temperature is reduced in dependence of the amount of melting, and we find the solution iteratively. 
The resulting temperature is prescribed as initial temperature, and the resulting melt fraction is prescribed as initial depletion. The porosity is assumed to be zero everywhere in the domain at the model start. 
We first let the model run in a low resolution of 1~km for 3~million years to produce a more realistic temperature and compositional structure that takes into account the dynamic effects of melt transport. Then we increase the resolution to 550~m throughout the model domain and 270~m within a distance of 7~km around the ridge axis, where melt is extracted from the domain. On this finer mesh, we compute another 3 million years of model evolution, which is approximately the time it takes for material to cross the distance from the ridge axis to the far end of the model domain. 
Finally, we let the model evolve for another 8000 years ($\sim$370 time steps) with a uniform cell size of 140~m. 
This allows us to export the final state of the model to data files and use them to create high-resolution initial conditions for the model runs presented in the following. The data files are freely available at \url{https://github.com/tjhei/paper-aspect-melt-paper-2-data} together with the input files and allow it to reproduce our results. 

\subsubsection{Influence of problem size}
\label{sec:problem-size}

To show that iteration numbers of the linear solver do not vary substantially with the size of the problem we are solving, we used the data files created from the final state of the 2-D mid-ocean ridge model described above to compute instantaneous 
flow models with different resolutions. Our results (see Table~\ref{table:2D-iteration-counts}) show that the number of GMRES iterations is insensitive to the problem size, and the number of Schur complement iterations that are done per GMRES iteration only increases slightly with problem size. 
This result highlights the usefulness of our new method for large-scale magma/mantle dynamics models. 

\begin{table}
  \centering
  \begin{tabular}{|c|rr|}
    \multicolumn{3}{c}{\textbf{Problem size: Number of linear solver iterations}}  \\
    \hline
    \#cells & GMRES iterations & average S block iterations \\ \hline
%    $1,536$ & 241 & 31462 & 131\\
    $6,144$ & 213 & 157\\
    $24,576$ & 176 & 199\\
    $98,304$ & 118 & 229\\
    $393,216$ & 118 & 261\\
    $1,572,864$ & 116 & 308\\
    $6,291,456$ & 119 & 343\\
    \hline
  \end{tabular}
  \caption{Iteration counts for a linear solver tolerance of $10^{-14}$.}
  \label{table:2D-iteration-counts}
\end{table}

\subsubsection{Influence of material properties}
\label{sec:material-properties}

\citet{sander_two_field} and \citet{sanderthreefield} have identified the ratio of compaction to shear viscosity as a key control on the rate of convergence of the iterative solver for the linear system we solve. Because the compaction viscosity is inversely proportional to the porosity, this ratio increases with decreasing porosity and becomes infinity in the limit of $\phi \rightarrow 0$ (which is the mathematically degenerate case) at the boundaries between regions with and without melt. 

As this boundary is present in most models of magma/mantle dynamics, and has the potential to slow down convergence of the linear solver substantially, we investigate the dependence of the convergence rate on the compaction-to-shear-viscosity ratio. 
In our new formulation, we address the part of the problem that relates to the interface between the solid and the partially molten region by rescaling the equation that contains the compaction viscosity, and introducing a threshold for the onset of two-phase flow.
Hence, in the following we will test the sensitivity of the iteration count to both the global compaction-to-shear-viscosity ratio 
and the choice of the melt transport threshold. 

For this purpose, we use the same setup as described above in section~\ref{sec:problem-size} to compute instantaneous 
flow models. 
When the compaction-to-shear-viscosity ratio $\xi/\eta$ is varied globally (Table~\ref{table:parameter-iteration-counts}), 
we see that there is a weak dependence of the GMRES iteration count on the compaction-to-shear-viscosity ratio, 
similarly to the results of \citet{sanderthreefield}. 
In addition, the S block iteration count increases with $\xi/\eta$. 
This is expected, as our formulation only addresses the increase of $\xi$ as  the porosity $\phi \rightarrow 0$.
However, this sensitivity to $\xi/\eta$ might not be problematic for realistic applications, as this ratio is expected to be on the order of 1--100 \citep{hewitt2008partial, takei2009viscous1, simpson2010multiscale, Katz2010, schmeling2012effective, alisic2014compaction}. 

Note that the values $\xi/\eta$ given in Table~\ref{table:parameter-iteration-counts} correspond to the ratio of the shear and compaction viscosity for a porosity $\phi=$~0.015. The actual ratio in the model varies by two orders of magnitude upwards from this reference value due to the different dependencies on porosity, which means that the ratio increases both for very low and very high porosities. 

\begin{table}
  \centering
  \begin{tabular}{|c|rr|}
    \multicolumn{3}{c}{\textbf{Compaction-to-shear-viscosity ratio: Number of linear solver iterations}}  \\
    \hline
    $\xi/\eta \, (\phi=1.5\%)$ & GMRES iterations & average S block iterations \\ \hline
    $2 \cdot 10^1$ & 74 & 116\\
    $2 \cdot 10^2$ & 124 & 147\\
    $2 \cdot 10^3$ & 124 & 248\\
    $2 \cdot 10^4$ & 125 & 345\\
    $2 \cdot 10^5$ & 175 & 403\\
    $2 \cdot 10^6$ & 182 & 434\\
    $2 \cdot 10^7$ & 183 & 435\\
    \hline
  \end{tabular}
  \caption{Iteration counts for a linear solver tolerance of $10^{-14}$, and 887939 Stokes degrees of freedom (98304 mesh cells).}
  \label{table:parameter-iteration-counts}
\end{table}

In addition, we also test the sensitivity of the solver convergence rate to the increase in the compaction-to-shear-viscosity ratio as  $\phi \rightarrow 0$ by varying the threshold for the onset of two-phase flow. 
The results (Table~\ref{table:threshold-iteration-counts}) reveal no sensitivity of the GMRES iteration count and only a very weak sensitivity of the S block iteration count to this threshold. 

\begin{table}
  \centering
  \begin{tabular}{|c|rr|}
    \multicolumn{3}{c}{\textbf{Threshold for melt transport: Number of linear solver iterations}}  \\
    \hline
    $K_\text{threshold}$ & GMRES iterations & average S block iterations \\ \hline
    $10^{-6}$ & 124 & 248\\
    $10^{-8}$ & 124 & 252\\
    $10^{-10}$ & 124 & 255\\
    $10^{-12}$ & 124 & 262\\
    $10^{-14}$ & 124 & 290\\
    \hline
  \end{tabular}
  \caption{Iteration counts for a linear solver tolerance of $10^{-14}$, and 887939 Stokes degrees of freedom (98304 mesh cells). }
  \label{table:threshold-iteration-counts}
\end{table}

Finally, we also want to provide a direct comparison to to the method of \cite{dannberg2016compressible}. 
Due to the strong dependence on problem size, we had to reduce the resolution, increase the threshold for the onset of two-phase flow and increase the solver tolerance of the model for this comparison, and we also removed the temperature dependence of viscosity. 
The results in Table~\ref{table:threshold-comparison} show both overall lower iteration counts and lower sensitivity to model parameters for the formulation developed in this study. They highlight that also for realistic application cases such as melt migration below mid-ocean ridges, our new method performs substantially better than the one developed in \cite{dannberg2016compressible}, 
and is feasible for accurately modelling the interface between regions with and without melt. 

\begin{table}
  \centering
  \begin{tabular}{|c|r|rr|rr|}
    \multicolumn{6}{c}{\textbf{Threshold for melt transport: Number of linear solver iterations}}  \\
    \hline
    \multicolumn{2}{|c|}{} & \multicolumn{2}{c|}{\textbf{\cite{dannberg2016compressible}}} & \multicolumn{2}{c|}{\textbf{this study}} \\
    \hline
    $K_\text{threshold}$ & $\phi_\text{threshold}$ & GMRES iterations & avg. S block iterations & GMRES iterations & avg. S block iterations \\ \hline
    $10^{0}$ & $10^{-2}$ & 1496 & 10 & 62 & 30\\
    $10^{-1}$ & $2.15 \cdot 10^{-3}$ & 3471 & 10 & 63 & 110\\
    $10^{-2}$ & $4.64 \cdot 10^{-4}$ & 12600 & 10 & 64 & 137\\
    $10^{-3}$ & $10^{-4}$ & 42272 & 10 & 64 & 166\\
    $10^{-4}$ & $2.15 \cdot 10^{-5}$ & 95869 & 10 & 64 & 166\\
    $10^{-5}$ & $4.64 \cdot 10^{-6}$ & -- & -- & 64 & 173\\
    \hline
  \end{tabular}
  \caption{Iteration counts for a linear solver tolerance of $10^{-8}$, and 62404 Stokes degrees of freedom (6144 mesh cells). 
  Entries marked with `--' indicate that there was no convergence reached after 100000 GMRES iterations.}
  \label{table:threshold-comparison}
\end{table}

\subsubsection{Scaling behaviour of the implemented solver}
\label{sec:scaling}

\begin{figure}
 \centering
 \includegraphics[width=\textwidth]{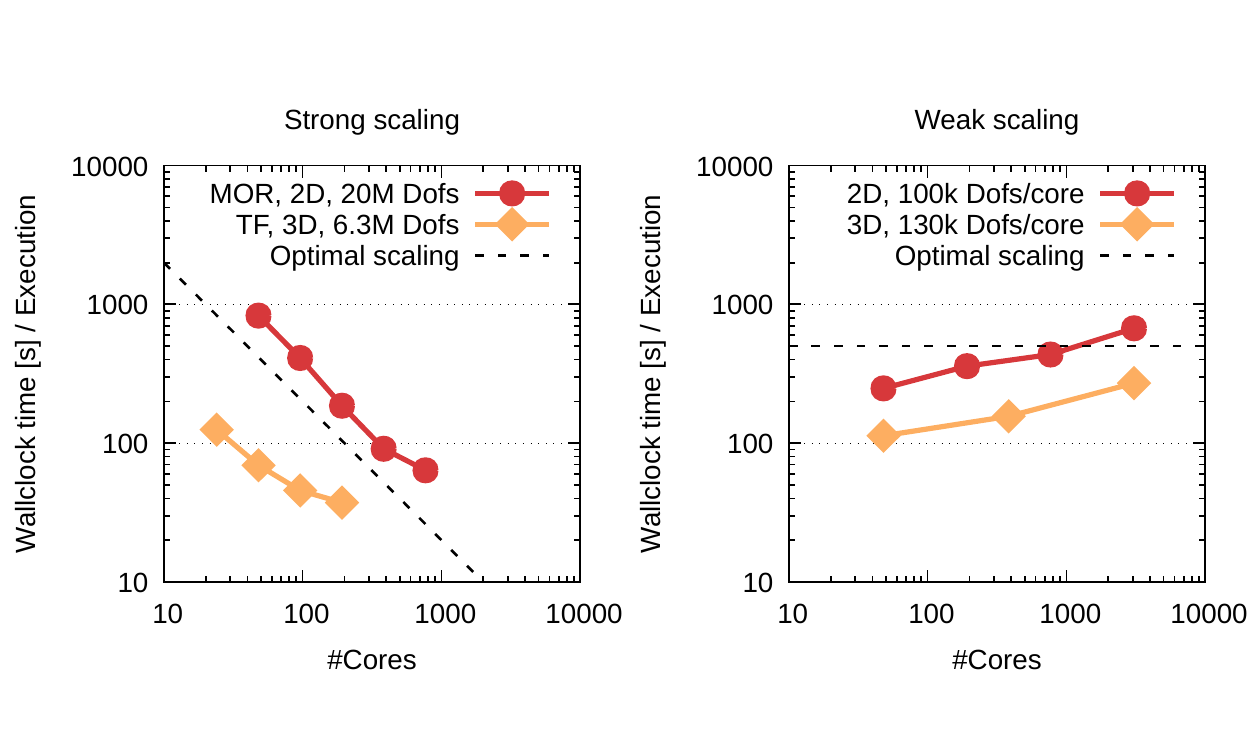}
 \caption{Strong and weak scaling results for the 2-D mid-ocean ridge (MOR) model described in Section~\ref{sec:2d-application} and the 3-D transform fault (TF) model described in Section~\ref{sec:3d-application}. The presented values represent the time required for solving the combined Stokes/Darcy equations once (i.e. without time spent for assembly, and equations for temperature, composition, and porosity). Note that the two model series use different iterative solver tolerances and values for $K_\text{threshold}$ so that absolute wallclock times cannot be compared between 2-D and 3-D. The scaling behaviour is not affected by these choices.}
  \label{fig:combined_scaling}
\end{figure}

In practice, not only the number of iterations, but also the wallclock time per iteration controls the computational cost of a model time step. Therefore we present scaling tests for the models of this section and Section~\ref{sec:3d-application} in Figure~\ref{fig:combined_scaling}. All scaling tests were done on Intel Xeon (Skylake) cores connected by an Intel Omnipath network at the Stampede 2 system of the Texas Advanced Computing Center (TACC).

Both models show a linear strong scaling to about 50,000 degrees of freedom (DoFs) per core (considering only solid velocity, fluid pressure, and compaction pressure DoFs); beyond that the efficiency drops significantly. The weak scaling results suggest a slightly less than optimal, but still acceptable scaling with model size, which leads to an increase of Stokes solver time by about a factor of 2.7 when increasing the model size by a factor of 64 (from 5 million DoFs to 327 million DoFs in 2-D, and from 6 million DoFs to 396 million DoFs in 3-D). These results are consistent with the slight increase in Schur complement iterations with model size discussed in Section~\ref{sec:problem-size} and show that our solver scales reasonably well to problem sizes of several hundred million and potentially a few billion degrees of freedom, although there is still room for optimization.

\subsubsection{A note on mesh refinement}
In \citet{dannberg2016compressible}, we discussed some strategies for adaptively refining the mesh in models 
with coupled magma/mantle dynamics. They mainly focused on refining the mesh based on solution variables or material properties. 
However, one can think of other useful mesh refinement strategies: 
One alternative is to just refine all cells where melt is present.
Another natural criterion that comes to mind is the intrinsic length scale of melt migration: the compaction length. 
The compaction length is defined as $\delta_c = \sqrt{(\xi + 4 \eta / 3) K_D}$
and is the length scale over which the compaction pressure
responds to variations in fluid flux \citep{spiegelman2007introduction, McK1984, spiegelman1993flow}. 
Hence, this length scale should be well resolved in numerical models that consider the compaction of partially molten rock. 
As the compaction length varies spatially and temporally, depending on the porosity of the rock and the material properties, adaptive mesh refinement can be a useful tool to make sure that the compaction length is resolved in an evolving model, while simultaneously saving computational resources by coarsening the mesh in regions with a larger compaction length. 

\begin{figure}
 \centering
 \includegraphics[width=0.45\textwidth]{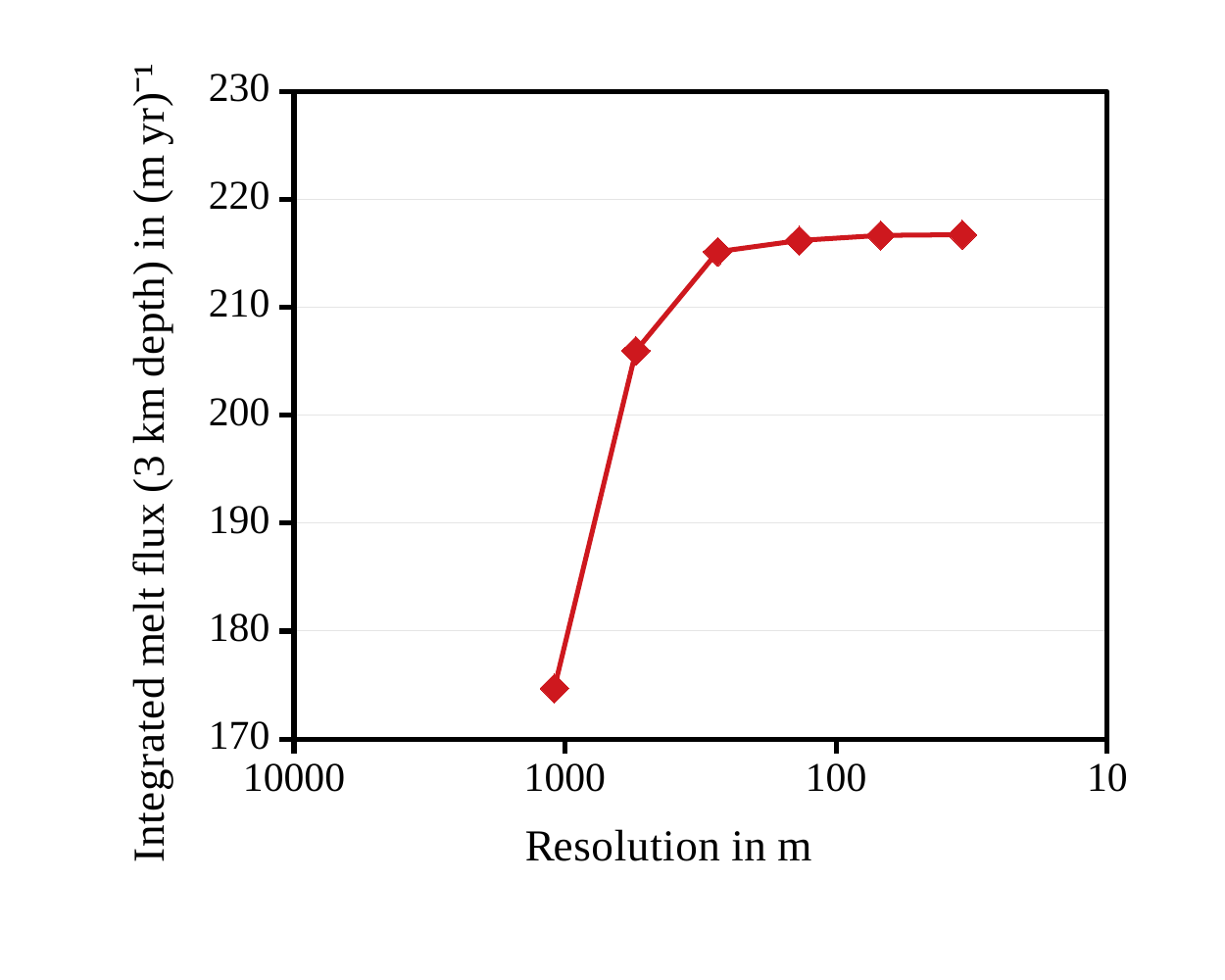}
 \includegraphics[width=0.45\textwidth]{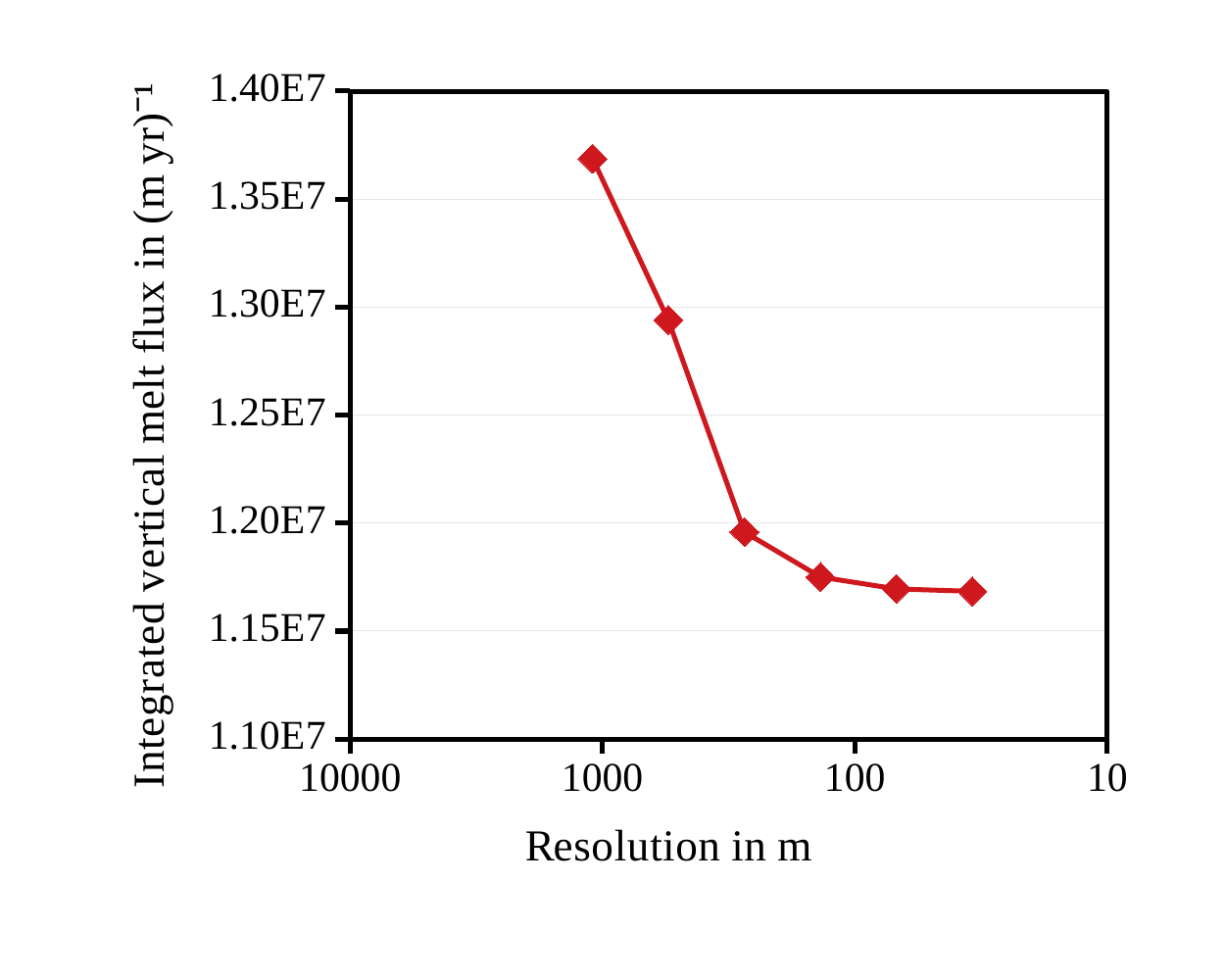}
 \includegraphics[width=0.45\textwidth]{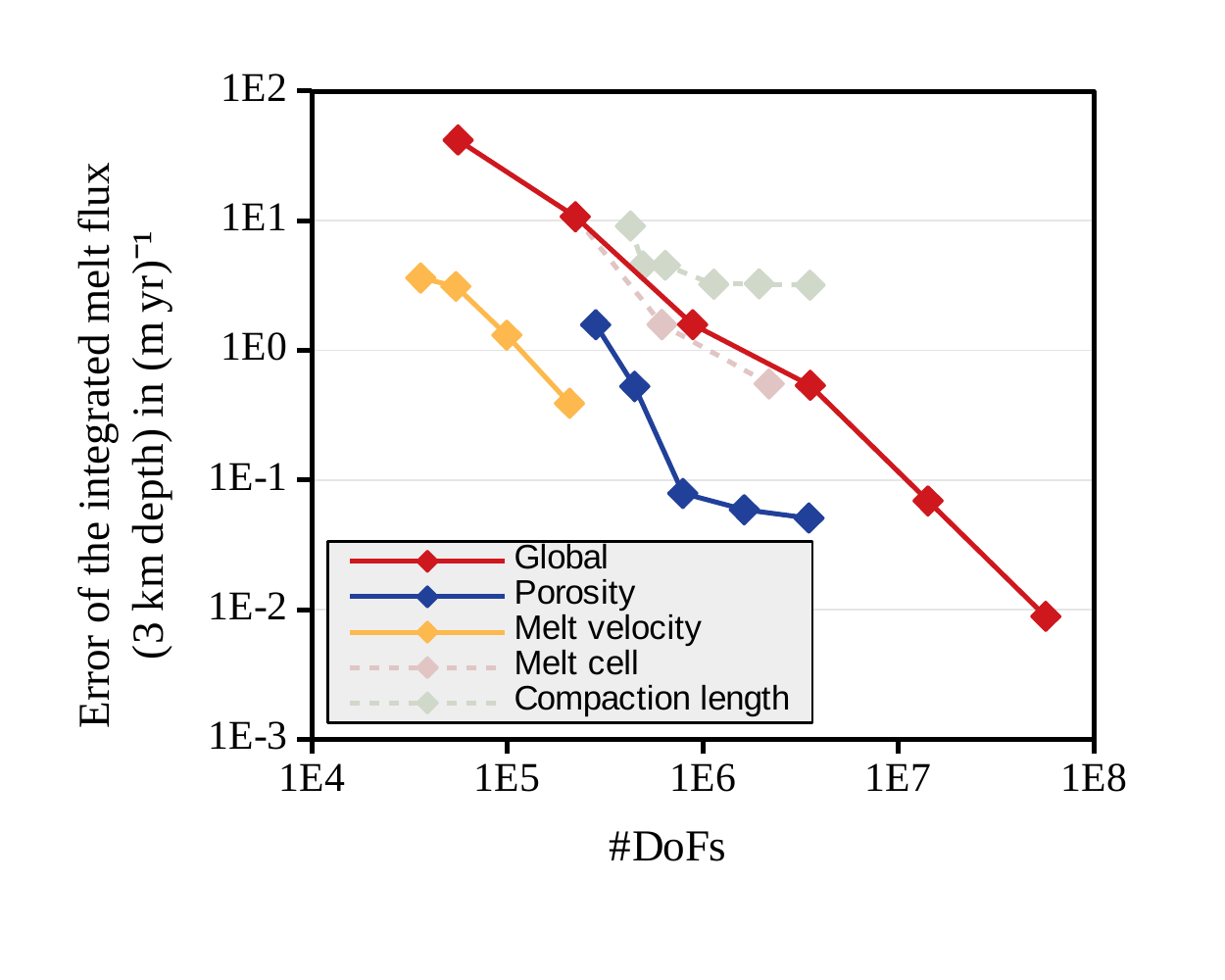}
 \includegraphics[width=0.45\textwidth]{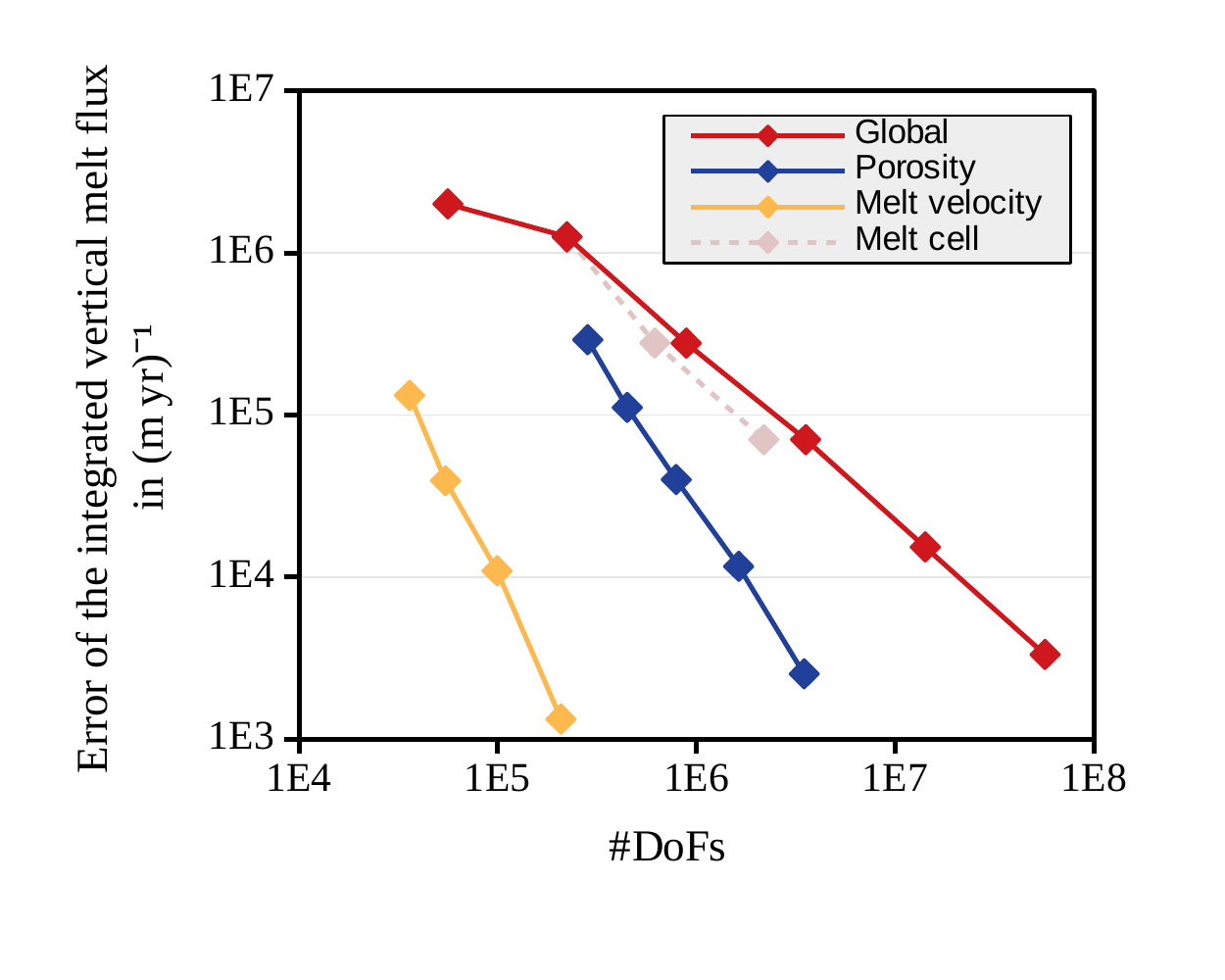}
 \caption{Melt flux in a 2-D mid-ocean ridge model for different resolutions. The top row shows the vertical melt flux integrated over a horizontal line in 3\,km depth (left) and the vertical melt flux integrated over the whole model domain (right) for models with uniform refinement of the mesh. 
 The bottom row panels feature the same quantities as the corresponding panels above, only that they show the error from the Richardson extrapolation of the data in the top row in logarithmic scale.
 The different data series represent uniform mesh refinement (red), and adaptive mesh refinement based on the porosity (blue) and the melt velocity (yellow), both using the Kelly error estimator, the presence of melt (light red) and the compaction length (light green).
 The results show that using adaptive mesh refinement can yield the same accuracy while using 1-2 orders of magnitude fewer degrees of freedom, and that for globally integrated quantities like the integrated melt flux, it can also yield a higher order of convergence. }
  \label{fig:adaptive_convergence}
\end{figure}

We implemented both mesh refinement strategies: 
One that refines all `melt cells', and one that adapts the size of the grid cells depending 
on the local compaction length, allowing it to define the minimum number of cells per compaction length
that should be present in the model. 
However, it becomes apparent that both of these strategies are inferior to refining based on solution variables, 
at least if the model output of interest is directly related to the solution variables (Figure~\ref{fig:adaptive_convergence}).  
Refining in `melt cells' performs slightly better than global refinement, but not nearly as good as refining based
on the porosity or the melt velocity, and using the compaction length as a refinement criterion is inferior even to
refining globally. 
The reason for that is that the compaction length decreases with increasing melt fraction, which means that the mesh is refined
first at the boundaries of the melting region. This increases the number of degrees of freedom, but does not accurately resolve the melt 
flux in regions where the porosity is large. 

The compaction length can still be a useful criterion to estimate the length scales of features emerging in a 
two-phase flow model, which can be used to set a minimum resolution in the partially molten regions.
In our mid-ocean ridge models, the compaction length (assuming a reference porosity of 0.5\%) is on the order of 10~km, 
which is well resolved in all models in Figure~\ref{fig:adaptive_convergence}, as the coarsest resolution is 2~km.
Nevertheless, locally, features may be substantially smaller than the compaction length, 
and our models require a global resolution of 140~m to reach an error of 1\%  for the global melt flux,  
which corresponds to $\sim$70 mesh cells per compaction length. 
This suggests that just resolving the compaction length might not be sufficient for accurately modelling of two-phase flow. 

\begin{figure}
 \centering
 \includegraphics[width=0.45\textwidth]{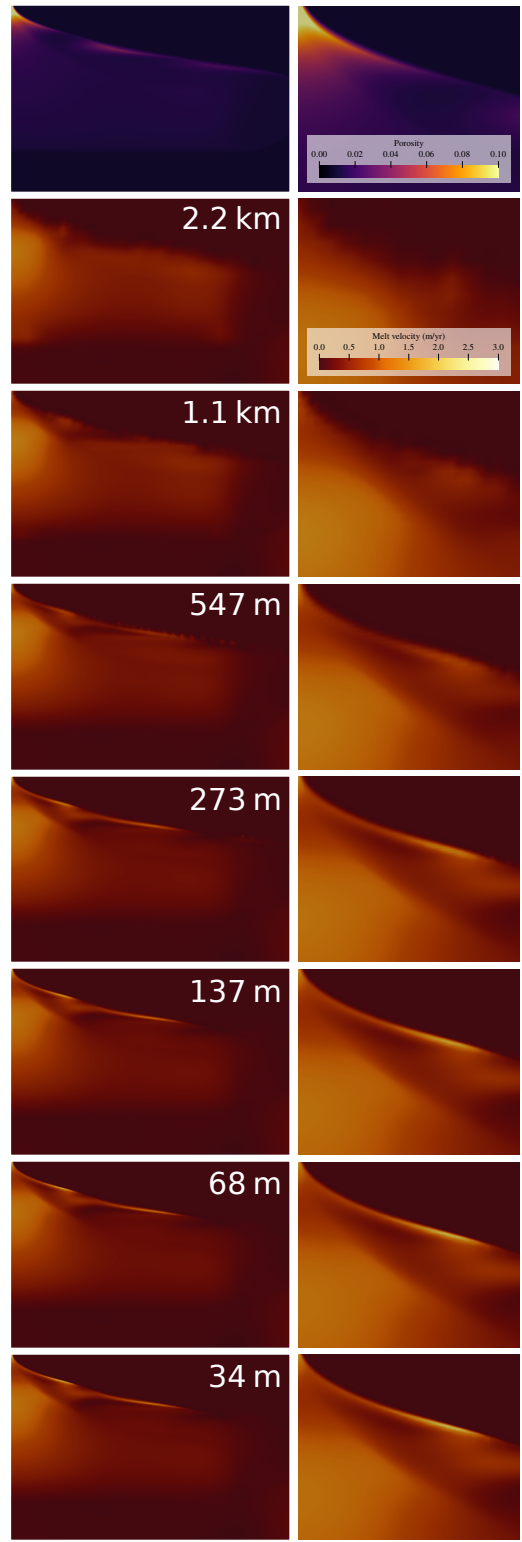}
 \caption{Porosity (top row) and melt velocity (all rows below) in a 2-D mid-ocean ridge model for different resolutions as given in Table~\ref{table:2D-iteration-counts}. The left column shows the whole model, the right column shows the part of the model closest to the ridge axis. Resolution increases from top to bottom, as specified by the white labels indicating the cell size in each model.}
  \label{fig:resolution}
\end{figure} 

\subsection{3-D Application: Oceanic transform fault}
\label{sec:3d-application}

To show the capability of our method to solve large-scale 3-D problems of coupled magma/mantle dynamics, 
we present an instantaneous mid-ocean ridge model that includes two ridge segments offset by a transform fault. 
We generated the initial conditions for this setup from the end state of the two-dimensional mid-ocean ridge 
model by mirroring the distribution of temperature, depletion and porosity with respect to the ridge axis and extending it uniformly in the third dimension, 
except for an offset of the ridge axis of 40~km in the center of the model. 
The material properties and boundary conditions are identical to the 2-D model described in Section~\ref{sec:2d-application}, and the new model boundaries at the front and back are free slip boundaries. 
The model extents are $170 \times 170 \times 70$~km, and we solve the (time-independent) coupled Stokes/Darcy equations on approximately 8.9 million cells (262 million degrees of freedom combined for solid velocity, fluid pressure and compaction pressure), 
as visualized in Figure~\ref{fig:melt_cells}. 
We make use of adaptive mesh refinement to increase the resolution in areas where melt is present, 
resulting in a cell size of approximately 550~m. 

\begin{figure}
 \centering
 \includegraphics[width=\textwidth]{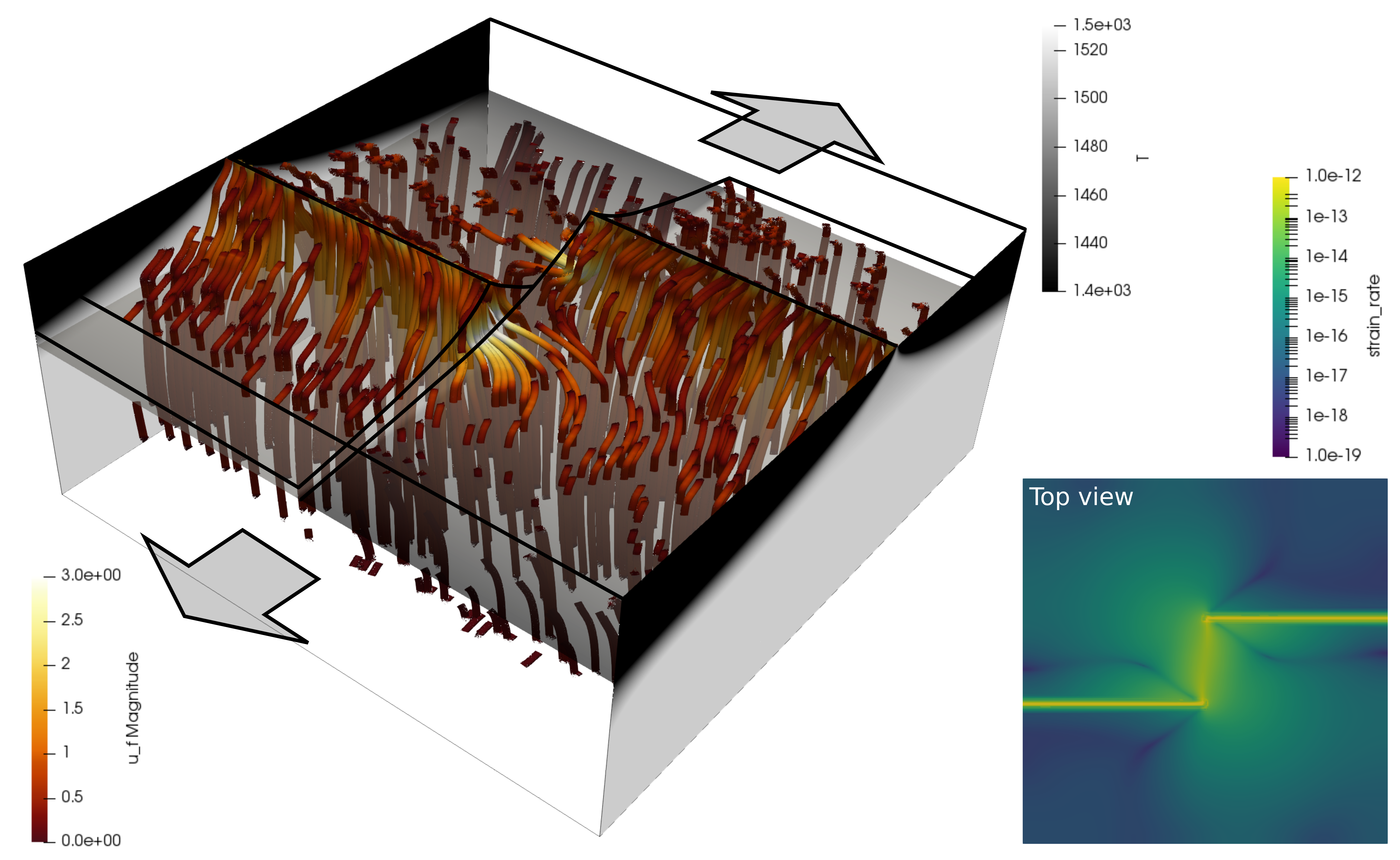}
 \caption{Visualization of a three-dimensional model of two mid-ocean ridge segments separated by a transform fault. Red-to-yellow streamlines show the melt velocity, highlighting the three-dimensional structure of the flow field (curved yellow--white streamlines cross the transform fault). Black-to-white background colours indicate temperature and grey arrows illustrate the prescribed spreading direction. The inset shows the deformation at the surface of the model.}
  \label{fig:transform_fault}
\end{figure}

The model output is illustrated in Figure~\ref{fig:transform_fault}, and shows that even though the temperature and porosity fields are symmetric with respect to the ridge axis of the individual ridge segments, the flow field evinces three-dimensional structures.
Melt is focused towards the ridge axis of the opposite ridge segment if that one is closer than the axis of the ridge segment the melt was generated at. Hence, melt crosses the transform fault, and the melt flux along the ridge axis decreases with increasing distance from the fault. 
In addition, deformation is not only localized at the two ridge segments, but the employed stress-limiter rheology
also leads to localization at the transform fault, where no melt is reaching the surface. 
It is clear that individual features of the flow field are likely to be different in a model with time evolution, 
where melt pathways are influenced by the acting stresses. However, our results highlight that high-resolution time-dependent three-dimensional models have a large potential to advance our understanding of the influence of transform faults and oblique spreading directions on the focusing of melt towards the ridge axis, and it is feasible to compute such models with the formulation we developed here. 

\section{Conclusions}
\label{sec:conclusions}

We have developed a new formulation of the governing equations of magma/mantle dynamics that allows it to accurately model the problem, even in the case of vanishing porosity and large ratios of compaction and shear viscosity. 
We achieve this by rescaling one of the solution variables, the compaction pressure, with the square root of the Darcy coefficient, 
and constraining the compaction pressure degrees of freedom to zero for very small porosities. 
This makes the linear system well-posed, even for small or vanishing porosities. 

Our numerical results show that the number of linear solver iterations is independent of the problem size, 
and that there is only a mild sensitivity to the model parameters. 
Hence, the method can be applied throughout a wide parameter range. 
Scaling tests reveal that our solver scales reasonably well to problem sizes of several hundred million, 
and potentially up to a few billion degrees of freedom. 
Most importantly, the solver convergence does not change with decreasing porosity, when the interface between solid and partially molten regions is approached. 

Finally, we demonstrated that our new formulation is suitable for modelling large-scale realistic problems of magma/mantle dynamics, 
such as melt generation and transport beneath mid-ocean ridges. 
Hence, we are confident that our new formulation and its implementation in the open source geodynamic modelling software \aspect{} 
will prove most valuable for exploring the the interactions of solid rock deformation and melt generation and transport in three dimensions. 

\section{Acknowledgements}
The authors would like to thank the Isaac Newton Institute for Mathematical Sciences, 
Cambridge, for support and hospitality during the programme Melt in the Mantle, 
which facilitated many fruitful discussions that lead to the idea for this paper.

All authors were partially supported by the Computational
Infrastructure for Geodynamics initiative (CIG), through the National
Science Foundation under Awards No.~EAR-0949446 and EAR-1550901,
administered by The University of
California-Davis.
TH was partially supported by the National Science Foundation awards
DMS-1522191, DMS-1820958, OAC-1835452 and by Technical Data Analysis,
Inc. through US Navy SBIR N16A-T003.
JD and RG were partially supported by the National Science
Foundation under award OCI-1148116 as part of the Software Infrastructure for
Sustained Innovation (SI2) program. 
JD gratefully acknowledges the support of the Deep Carbon Observatory.
The authors acknowledge the Texas Advanced Computing Center (TACC, \url{http://www.tacc.utexas.edu }) at 
The University of Texas at Austin for providing high-performance computing 
resources that have contributed to the research results reported within this 
paper. 
Clemson University is acknowledged for generous allotment of compute
time on the Palmetto cluster.

The authors greatly appreciate all of these sources of support.

\bibliographystyle{gji}
\bibliography{paper}

\end{document}